# Xona Pulsar Single-Satellite Positioning: System Perspective and Experimental Validation

Thyagaraja Marathe, Tyler G. R. Reid, Srinivas Tantry, Michael O'Meara
*Xona*

## BIOGRAPHIES

**Dr. Thyagaraja Marathe** is a Staff Research Engineer at Xona, focusing on PNT with LEO satellite constellations. He earned a PhD in Geomatics Engineering (2016) from the University of Calgary, where he also completed postdoctoral research, an M.S. (2011) from BITS Pilani, and a B.Eng. (2005) from VTU, India. His background spans advanced GNSS technologies, including precise positioning, corrections, and receiver innovation, and he has previously worked at Rx Networks and Accord Software.

**Dr. Tyler G. R. Reid** is co-founder and CTO of Xona. Previously, Tyler worked as a Research Engineer at the Ford Motor Company in localization and mapping for self-driving cars. He has also worked as a software engineer at Google and as a lecturer at Stanford University, where he co-taught the GPS course. Dr. Reid received his PhD (2017) and MSc (2012) in Aeronautics and Astronautics from Stanford, where he worked in the GPS Lab.

**Srinivas Tantry** is a GNSS Engineer at Xona where he focuses on evaluating positioning and timing performance of Xona's Pulsar satellite navigation system. He has previously worked at Rx Networks on GNSS corrections and at Accord Software & Systems on GNSS receiver design and development. He received his MEng in Geomatics Engineering from the University of Calgary.

**Michael O'Meara** is a GNSS Systems Specialist at Xona, performing various duties such as verification and validation of hardware and software, as well as technical regulatory analysis. Michael has previously worked for Magnestar Inc. developing a software-as-a-service for regulatory SATCOM simulations. He obtained a BEng in Aerospace Engineering from Concordia University.

## Abstract

Xona is deploying Pulsar, a low Earth orbit (LEO) commercial satellite navigation system designed to deliver resilient positioning, navigation, and timing (PNT) in environments where traditional solutions fall short. Pulsar satellites broadcast dedicated navigation signals optimized for commercial users. This brings rapid geometry change, strong Doppler observability, and robust timing, enabling new approaches to positioning even when only one satellite is visible.

Internet of Things (IoT) is an application where availability matters more than sub-meter accuracy, such as operations in urban canyons, semi-indoor spaces, or other constrained environments. Many of these applications involve platforms with strict size, weight, and power (SWaP) limitations, including battery-powered systems that cannot accommodate complex multi-sensor architectures. By leveraging the inherent dynamics and signal strength of LEO satellites, Pulsar offers a pathway to maintain navigation capability under these challenging conditions without requiring specialized user hardware.

Here we present a single-satellite positioning (SSP) concept that uses all available measurements from Pulsar signals to estimate user position and receiver clock states without external aiding. In early phases of Pulsar deployment, there will be only one or two satellites in view, but this still brings value to stationary or near-stationary users. Along with the description of algorithmic details, there is emphasis on the system-level implications. SSP not only enables positioning when only one Pulsar satellite is visible, but also reduces reliance on dense constellations, and supports integration into resource-constrained platforms.

We present both simulation and live sky testing results. First, simulations using a high-fidelity constellation simulator configured for Pulsar signals provide controlled performance assessment across diverse geometries and environmental conditions. Second, we include early findings from a Pulsar-enabled receiver, incorporating real observations from the Pulsar-0 satellite on orbit. Preliminary live testing of Pulsar-0 signals demonstrates achievable position accuracies at the meter level both outdoors and indoors, highlighting the system's potential under varying reception conditions.

# 1. INTRODUCTION

LEO constellations are emerging as a complementary layer to existing MEO GNSS. These can offer higher received signal power and rapid geometry evolution that improve availability, resilience, Doppler observability, and time to fix, particularly in urban canyons and semi-indoor environments (Baron, Gurfil, & Rotstein, 2024; Knight, 2025; Çelikbilek & Lohan, 2024). Several studies highlight the inherent advantages of LEO PNT: stronger signals (tens of dB above typical GNSS at the surface), faster angular rates that drive informative Doppler and Doppler rate, and high satellite counts for layered robustness (Fabra, Egea–Roca, López–Salcedo, & Seco–Granados, 2024). Xona's Pulsar is a dedicated commercial LEO PNT system broadcasting GNSS-compatible L-band signals with higher power and security features, designed to interoperate with existing receiver ecosystems (Reid, et al., 2025). The production-class satellite, Pulsar-0, is now undergoing in-orbit testing ahead of constellation build-out; the stated goals include stronger signals, authentication, and centimeter-level positioning in conjunction with corrections services (Khalil, 2025; Leclère, Marathe, & Reid, 2025).

LEO's fast geometry change enables single-satellite positioning for stationary or semi-stationary users by stacking measurements over time. The concept is supported by observability analyses showing that position and clock states are theoretically observable from a single LEO satellite given known ephemeris; practical performance depends on pass duration, Doppler precision, and modeling fidelity (Sabbagh & Kassas, 2023). The Transit system, a navigation system in LEO orbit and which was based on Doppler observations, was available for military use since 1963. Observations collected over a single 15-minute pass provided a user position precision of 20 to 40 m (Black, Jenkins, & Pryor, 1976). This system was not designed to enable generation of accurate pseudoranges; therefore, improving accuracy beyond a limit using data from a single pass was not possible. A growing body of experimental and simulation results demonstrates the utility of joint pseudorange + Doppler estimation with LEO satellites (signals of opportunity (SOP) and dedicated PNT): joint formulations improve accuracy and availability relative to pseudorange-only, while sensitivity analyses underline the central role of Doppler error and satellite ephemeris/velocity accuracy (Headlee, Lo, & Walter, 2025; Xu, Liu, Lei, Fang, & Jiang, 2025; Morichi, Minetto, Nardin, Zocca, & Dovis, 2024). LEO altitudes shorten path loss and produce large Doppler and Doppler rate, offering benefits for acquisition, tracking, and range-rate observability. These dynamics enable SSP for static users by exploiting line-of-sight sweep across a pass.

This work focuses on a single-satellite positioning paradigm that leverages time and spatial diversity from one or multiple LEO passes. The approach jointly processes pseudorange and Doppler to estimate user ECEF position and receiver clock bias/drift without external aiding. It is particularly relevant to early deployment phases and availability-centric applications. This paper presents a system-level perspective for Xona Pulsar SSP, articulates a standard mathematical formulation for batch estimation from pseudorange and Doppler, and validates feasibility through simulations and early live sky experiments.

Section 2 presents an overview of the Pulsar constellation and signals. In Section 3, we provide a description of single satellite positioning paradigm, and methodology is given in Section 4. Section 5 provides a discussion on experimental results. Section 0 outlines a summary of conclusions from experiments. Section 7 provides a use-case overview and possible enhancements along with future work.

# 2. XONA PULSAR OVERVIEW

Pulsar targets high-accuracy and resilient PNT via a LEO architecture and modernized signals intended to interoperate with existing GNSS ecosystems. Pulsar is a commercial 258 satellite LEO constellation planned by Xona, specifically designed for PNT. The Pulsar in-orbit validation (IOV) satellite, Pulsar-0, was launched on June 23, 2025, and is currently in operation after successful commissioning.

Pulsar shares several fundamental characteristics with existing GNSS systems. Like GNSS, Pulsar satellites orbit the Earth and transmit RF signals that enable users to compute satellite positions and derive pseudoranges for PVT solutions. The signals are broadcast in the L1 and L5 frequency bands, close to existing GNSS allocations, allowing compatibility with current receiver RF chains. Pulsar employs direct sequence spread spectrum (DSSS) with primary and secondary PRN codes, supporting both pilot and data channels. One signal uses binary phase shift keying (BPSK) modulation.

In addition to these similarities, Pulsar introduces advanced features that enhance performance and flexibility. Operating in LEO significantly reduces transmission distance, resulting in much higher received power. Pulsar adopts bandwidth-efficient enhanced Feher's QPSK (EFQPSK) and code shift keying (CSK) modulation, enabling higher data rates and faster time-to-first-fix (TTFF). Its flexible data structure supports future evolutions with backward compatibility, while advanced capabilities such as satellite-based GNSS corrections, data encryption, authentication, and range integrity further distinguish Pulsar from traditional GNSS.

**2.1. Constellation**

To achieve Earth coverage comparable to a medium Earth orbit (MEO) GNSS constellation, LEO systems require a substantially larger number of satellites (Reid, 2017). The planned Pulsar constellation will consist of 258 satellites distributed across 18 orbital planes. Two plane types are employed to ensure global coverage: inclined planes with an inclination of 53° and polar planes with an inclination of 97°. Satellites within each plane are uniformly spaced, and the relative phasing between adjacent planes is fixed to maintain an even distribution across the entire constellation.

A nominal altitude of 1080 km was chosen to balance the number of satellites required with the achievable dilution of precision (DOP) (Reid, et al., 2020). In contrast, the Pulsar-0 satellite operates at an altitude of approximately 520 km with a 97° inclination. At this lower altitude, satellites complete an orbit in less than two hours compared to roughly 12 hours for GPS, resulting in significantly higher relative motion. Consequently, Doppler shifts observed on Earth are about eight times greater for Pulsar than for GPS, and the effect is even more pronounced for higher-order derivatives: Doppler rate is roughly 250 times higher for the Pulsar full operational capability (FOC) constellation and about 500 times higher for Pulsar-0. These dynamics impose important considerations for receiver design and performance.

Table 1 provides a summary of the main orbital parameters for the nominal constellations (Xona Space Systems, 2025; Leclère, Marathe, & Reid, 2025).

**Table 1:** Nominal Pulsar IOV, Pulsar FOC and GPS Constellations Characteristics

| Parameters | Pulsar IOV | Pulsar FOC Polar Orbit | Pulsar FOC Inclined Orbit | GPS |
|---|---|---|---|---|
| Number of orbital planes | 1 | 6 | 12 | 6 |
| Orbital planes RAAN offset | - | 60° (360° / 6) | 30° (360° / 12) | 60° (360° / 6) |
| Number of satellites per plane | 1 | 11 | 16 | 4 |
| Number of satellites | 1 | 66 | 192 | 24 |
| Orbit inclination | 97° | 97° | 53° | 55° |
| Orbit eccentricity | 0 | 0 | 0 | 0 |
| Orbit altitude | 520 km | 1080 km | | 20180 km |
| Orbit radius | 6891 km | 7451 km | | 26551 km |
| Orbital period | 95 min ≈ 1.6 h | 107 min ≈ 1.8 h | | 718 min ≈ 12.0 h |
| Satellite orbital speed | 7606 m/s | 7314 m/s | | 3875 m/s |
| | 27380 km/h | 26331 km/h | | 13949 km/h |
| Maximum ECEF satellite speed | 7683 m/s | 7400 m/s | 7001 m/s | 3187 m/s |
| | 27659 km/h | 26640 km/h | 25202 km/h | 11472 km/h |
| Maximum relative speed to a static user | 7103 m/s | 6327 m/s | 5986 m/s | 765 m/s |
| | 25572 km/h | 22779 km/h | 21549 km/h | 2753 km/h |
| Max. L1 band carrier Doppler | 37752 Hz | 33629 Hz | 31813 Hz | 4018 Hz |
| Max. L5 band carrier Doppler | 28208 Hz | 25127 Hz | 23771 Hz | 3001 Hz |

## 2.2. Signals

Pulsar follows the same fundamental principles as legacy GNSS, and its signal design shares many similarities with existing systems. The constellation broadcasts two signals: X1 in the L1 band and X5 in the L5 band. Each signal includes a carrier, a primary PRN code, an overlay code, and data bits transmitted using different modulation schemes, along with distinct pilot and data channels. The detailed characteristics of Pulsar and GPS signals in these frequency bands are summarized in Table 2 (Xona Space Systems, 2025; Leclère, Marathe, & Reid, 2025).

**Table 2:** Pulsar and GPS Signals Characteristics

| Parameters | L1 band | | L5 band | |
|---|---|---|---|---|
| | Pulsar X1 | GPS L1 C/A | Pulsar X5 | GPS L5 |
| Carrier frequency | 1593.3225 MHz (155.75 × 10.23) | 1575.42 MHz (154 × 10.23) | 1190.51625 MHz (116.375 × 10.23) | 1176.45 MHz (115 × 10.23) |
| Bandwidth* | 1.77 MHz | 2.046 MHz | 17.7 MHz | 20.46 MHz |
| Modulation | EFQPSK | BPSK | EFQPSK + CSK | QPSK |
| Minimum received power** | −148.2 dBW | −158.5 dBW | −144.9 dBW | −154.0 dBW |
| Maximum received power | −139.1 dBW | −153.0 dBW | −136.2 dBW | −150.0 dBW |
| Data/pilot channel combination | Quadrature (I/Q) | - | Quadrature (I/Q) | |
| Data/pilot channel power split | 50 % / 50 % | - | 50 % / 50 % | |
| Primary PRN code family | Kasami (small set) | Gold | Extended Gold | |
| Primary PRN code chip rate | 1.023 Mchip/s | | 10.23 Mchip/s | |
| Primary PRN code length | 1023 chip = 1 ms | | 10230 chip = 1 ms | |
| Overlay code chip rate | 1000 chip/s | - | 1000 chip/s | |
| Overlay code length (pilot) | 100 chip = 100 ms | - | 100 chip = 100 ms | 10 chip = 10 ms |
| Overlay code length (data) | - | - | - | 20 chip = 20 ms |
| Symbol duration (data) | 1 ms | 20 ms | 2 ms | 10 ms |
| Symbol rate (data) | 1000 symbol/s | 50 symbol/s | 500 symbol/s | 100 symbol/s |
| Bit rate (data) | 1000 bit/s | 50 bit/s | 4000 bit/s | 100 bit/s |

\* Indicated bandwidths contain 99.5 % of the total power for Pulsar and correspond to the main lobe for GPS.
\** Minimum received powers are specified for an elevation angle of 10° or more for Pulsar and 5° or more for GPS.

## 3. SINGLE SATELLITE POSITIONING PARADIGM

Single-satellite positioning leverages the unique dynamics of LEO satellites to estimate user position from a single spacecraft pass. The basic principle relies on the high angular velocity of LEO satellites, which produces a distinctive Doppler frequency profile as observed by the receiver. This profile encodes longitudinal information because the Doppler curve shape depends on the ground track relative to the user. Additionally, the point where the Doppler shift transitions from positive to negative corresponds to the closest approach of the satellite, providing an estimate of latitude. By accumulating Doppler measurements over time and combining them with known satellite ephemerides, SSP converts temporal diversity into spatial diversity, enabling position estimation even with one satellite pass. This approach is feasible in LEO because the rapid motion and strong Doppler signatures provide sufficient observability within a short time window; in contrast, MEO satellites move slowly, producing weak Doppler variations and lack observability that make SSP impractical for GNSS systems like GPS or Galileo (Shi, Zhang, & Li, 2023; Leclère, Marathe, & Reid, 2025).

In Pulsar's case, SSP performance is further enhanced by combining carrier Doppler with pseudorange measurements. Pseudoranges provide absolute range constraints, while Doppler adds velocity-dependent information, improving geometry and reducing ambiguity in position estimation. This fusion is possible because Pulsar is a dedicated PNT system with well-defined signal structures and precise timing, allowing accurate extraction of both observables (Leclère, Marathe, & Reid, 2025; Morichi, Minetto, Nardin, Zocca, & Dovis, 2024). Conversely, opportunistic use of non-PNT LEO payloads (e.g., broadband satellites) faces significant limitations: lack of synchronized clocks, incomplete signal specifications, and uncertain ephemerides degrade pseudorange accuracy, leaving Doppler-only solutions with poor dilution of precision and sensitivity to user clock drift. Differential techniques can partially mitigate these errors, but single-satellite opportunistic positioning remains coarse and unsuitable for high-integrity applications (Headlee, Lo, & Walter, 2025).

We consider a stationary (or slow) user illuminated by a single LEO satellite. Over a time-window covering part of a pass, the receiver stacks code-based pseudorange and carrier-Doppler (converted to range-rate) to estimate position and receiver clock states. Observability analyses show that for a stationary user, position and time are theoretically observable from a single LEO satellite with known ephemeris; practical performance depends on pass length, measurement noise, and modeling fidelity (Sabbagh & Kassas, 2023). The objective is not decimeter precision but robust availability with meter/tens-of-meters accuracy depending on conditions—accuracy that is sufficient for many IoT and logistics applications (Headlee, Lo, & Walter, 2025; Xu, Liu, Lei, Fang, & Jiang, 2025).

## 4. METHODOLOGY

A description of the measurement models and the estimation strategies are given in this section.

### 4.1. Measurement models and state:

The state vector for a static receiver is given as

$$\mathbf{x} = \begin{bmatrix} \mathbf{r} \\ B \\ d \end{bmatrix} = [x \quad y \quad z \quad B \quad d]^{\mathrm{T}}, \tag{1}$$

where $\mathbf{r} \in \mathbb{R}^3$ [m] is receiver position, $B$ [m] is receiver clock bias, and $d$ [m/s] is receiver clock drift, and $[\ ]^{\mathrm{T}}$ denotes the transpose.

The geometric range is

$$\rho_i^{geo} = \|\mathbf{s}_i - \mathbf{r}\|, \tag{2}$$

and the line-of-sight unit vector is

$$\mathbf{u}_i = \frac{(\mathbf{s}_i - \mathbf{r})}{\rho_i^{geo}}, \tag{3}$$

with $(\mathbf{s}_i, \mathbf{v}_i)$ the satellite state (position and velocity)

The pseudorange model for code measurements is,

$$\rho_i = \rho_i^{geo} + B + d\Delta t_i + \varepsilon_{\rho,i},$$
$$\Delta t_i = t_i - t_0 \tag{4}$$

The Doppler to range-rate conversion is given as,

$$\dot{\rho}_i^{obs} = -\lambda D_i \tag{5}$$

and the range-rate model as,

$$\dot{\rho}_i = \mathbf{u}_i^T \mathbf{v}_i,$$
$$\dot{\rho}_i^{obs} = \dot{\rho}_i + d + \varepsilon_{D,i} \tag{6}$$

Formulation for the residuals and Jacobians is as given below.

Residual at epoch $i$:

$$\mathbf{r}_i = \begin{bmatrix} \rho_i - \rho_i^{geo} - B - d\Delta t_i \\ \dot{\rho}_i^{obs} - \dot{\rho}_i - d \end{bmatrix} \quad (7)$$

Jacobian at epoch $i$:

$$\mathbf{J}_i = \begin{bmatrix} -\mathbf{u}_i^T & 1 & \Delta t_i \\ \left(\dfrac{\mathbf{v}_i - (\mathbf{v}_i^T \mathbf{u}_i)\mathbf{u}_i}{\rho_i^{geo}}\right)^T & 0 & 1 \end{bmatrix} \quad (8)$$

Stacking all $N$ epochs gives $[\mathbf{r}_1^T \cdots \mathbf{r}_N^T]^T$ and $\mathbf{J} = \begin{bmatrix} \mathbf{J}_1 \\ \mathbf{J}_2 \\ \vdots \\ \mathbf{J}_N \end{bmatrix}$. (Derivatives follow standard GNSS Jacobian forms for range and range-rate.)

The estimation strategies used in this work are given below:

Gauss-Newton (GN):

$$\Delta x = (\mathbf{J}^T \mathbf{J})^{-1} \mathbf{J}^T \mathbf{r}. \quad (9)$$

Weighted GN / Levenberg–Marquardt (LM):

$$(\mathbf{J}^T \mathbf{W} \mathbf{J} + \mu \mathbf{I})\Delta \mathbf{x} = \mathbf{J}^T \mathbf{W} \mathbf{r}, \quad (10)$$

with $\mathbf{W} = \text{diag}(\sigma_\rho^{-2}, \sigma_D^{-2}, \cdots)$ and adaptive damping $\mu$.

A stacked nonlinear least squares problem over a pass window (single pass) or over multiple passes (multi-pass) is then solved.

### 4.2. Choice of Estimation Strategy for Different Data Sources

In this work, we adopt an unweighted Gauss–Newton (GN) approach for software-based simulations and a weighted Levenberg–Marquardt (LM) method for live-sky data processing. The rationale is twofold: (i) simulations provide idealized, low-noise conditions and well-conditioned geometry, where GN offers fast convergence with minimal computational overhead; (ii) real-world data introduces measurement noise, outliers, and potential ill-conditioning, making LM's damping and weighting essential for numerical stability and robustness. Despite these differences, the core formulation—joint estimation of receiver position and clock states from pseudorange and Doppler over a batch window—remains identical across both strategies, ensuring methodological consistency.

### 4.3. Extensions for multi pass processing

The original method was designed for a single LEO satellite pass, estimating position and receiver clock parameters using pseudorange and Doppler observations. To adapt this approach for multi-pass data, the algorithm was modified so that the position remains common across all passes, while clock bias and drift are estimated separately for each pass. This required expanding the state vector from five parameters, i.e., three for position and two for clock terms, to include additional bias and drift pairs for each pass, i.e., expanding the state vector from $[x, y, z, B, d]$ to $[x, y, z, B_1, d_1, B_2, d_2, \cdots]$ for multiple passes. The Jacobian and residual computation were also updated so that each observation only influences its corresponding pass-specific clock parameters, while still contributing to the shared position estimate. All observations from multiple passes are then combined in a single least-squares adjustment, preserving the original solver structure and iteration logic.

This modification brings significant benefits to the SSP paradigm. By leveraging multiple passes, the algorithm improves geometric diversity, which reduces dilution of precision and enhances position accuracy compared to using a single short arc. It makes SSP more practical for applications where continuous visibility is not possible, such as low Earth orbit constellations or sparse tracking scenarios. However, the approach has limitations: if one pass is much noisier than others, its residuals can degrade the combined solution unless adaptive weighting or robust estimation techniques are applied. Additionally, the method assumes independent clock states for each pass, which is valid for large time gaps.

## 5. EXPERIMENTAL RESULTS

Proposed methods were tested in two different domains, namely, software simulations and Pulsar-0 navigation pass data trials. For the first case, a commercial Pulsar-enabled GNSS simulator was used to simulate Pulsar constellation and data containing the required observations and satellite information was logged from the simulator runs. In the second mode, a Pulsar-enabled receiver was used in the field to collect actual satellite data from the navigation passes from Pulsar-0, which is the in-orbit validation satellite from Xona. Only observables on X1 frequency are used in this work.

### 5.1. Software Simulations

A short description of the configuration used for software simulation is provided herein. A commercial Pulsar-enabled GNSS simulator was used to generate CSV log files containing parameters such as time, satellite positions, pseudorange, and Doppler for each satellite. Several GNSS simulators are currently supporting Pulsar (Safran, 2024; Spirent, 2024). For this study, Safran's Skydel GNSS Simulator Software was used.

Since the Pulsar constellation is not yet fully deployed, the simulator initially used default parameters based on the nominal constellation, with some randomness added. Using nominal constellations gives characteristics very similar to those that would be obtained with the actual constellation in space.
Here are the common parameters of the simulations:
- The duration of the simulation is 3 days, because it is long enough to capture unbiased statistics and metrics.
- Since user latitude is one factor that could impact the observations, the analysis was performed at three latitudes to ensure different coverage: 0°, 30°, 60°.
- The longitude is 0° and the altitude is 0 m. The choice of longitude would not significantly matter since statistics were computed considering full constellation comprising of 258 satellites.
- An elevation mask of 0° is considered.

There are about 6 to 7 satellite passes per day for each satellite. Considering a 3-day log for 258 satellites, there are more than 4600 individual satellite passes in the logged data. These satellite passes are processed individually to obtain single pass-based statistics and observations from consecutive passes are combined to obtain position statistics that combined data from two and three passes.

Figure 1 illustrates the horizontal positioning accuracy, achieved when processing observations from one, two, and three passes across different user latitudes. The cumulative distribution of errors reveals that even a single-pass solution can deliver positioning accuracy on the order of a few tens of meters, which is sufficient for many availability-driven applications. However, occasional error excursions into hundreds of meters due to the lack of observability highlight the limitations of relying on a single pass, particularly under challenging geometries or signal conditions. By combining observations from multiple passes, the solution benefits from improved geometry and time diversity, which significantly enhances both accuracy and error consistency. This multi-pass approach reduces outlier occurrences and tightens the error distribution, enabling more reliable positioning performance across diverse environments.

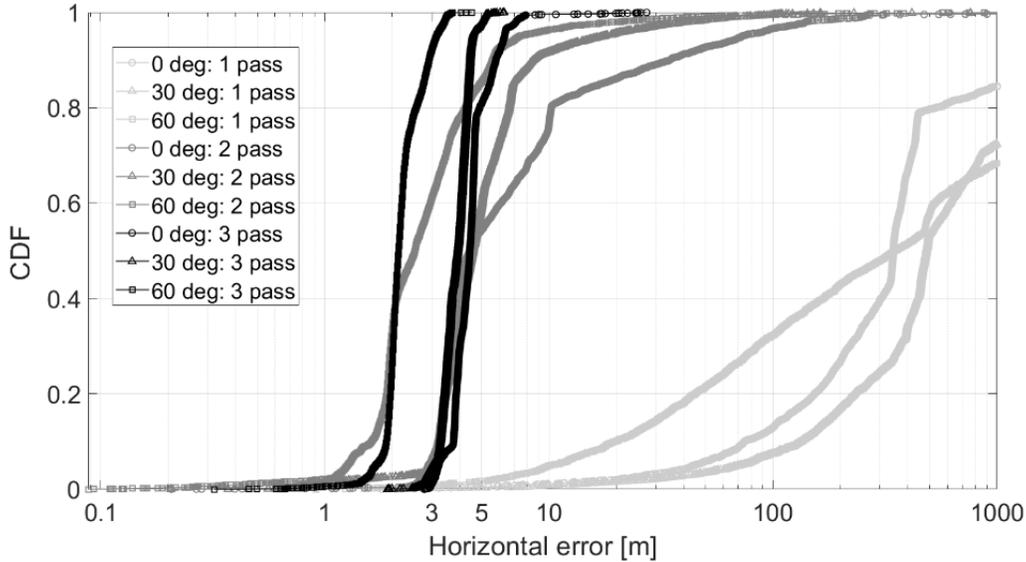

**Figure 1:** Software simulation: Horizontal position accuracy for one, two and three passes for 0°, 30°, 60° latitudes

A summary of statistics of horizontal errors for the simulation-based results is given in Table 3.

**Table 3:** SSP horizontal user position accuracy software simulations

| User Latitude | Configuration | 1-sigma Horizontal error [m] | 2-sigma Horizontal error [m] |
|---|---|---|---|
| **0°** | 1-pass | 404 | > 1000 |
|  | 2-pass | 8 | 72 |
|  | 3-pass | 4.6 | 6.3 |
| **30°** | 1-pass | 800 | > 1000 |
|  | 2-pass | 5.5 | 18 |
|  | 3-pass | 4.2 | 4.6 |
| **60°** | 1-pass | 956 | > 1000 |
|  | 2-pass | 3.4 | 7 |
|  | 3-pass | 2.3 | 3.2 |

### 5.2. Early live-sky trials

Using a prototype Pulsar-enabled receiver, we collected both pseudorange and Doppler measurements during navigation passes to evaluate positioning performance. Data was acquired using a combination of outdoor and semi-indoor antenna setups during a high-elevation satellite pass over Montreal on September 27, 2025. This configuration allowed us to assess performance across varying signal environments, highlighting the robustness of Doppler-aided solutions even in partially attenuated conditions. The prototype receiver used herein is a standard sensitivity receiver and not specifically tuned for high sensitivity use case. A representation given in Figure 2 provides some details about the data collection site.

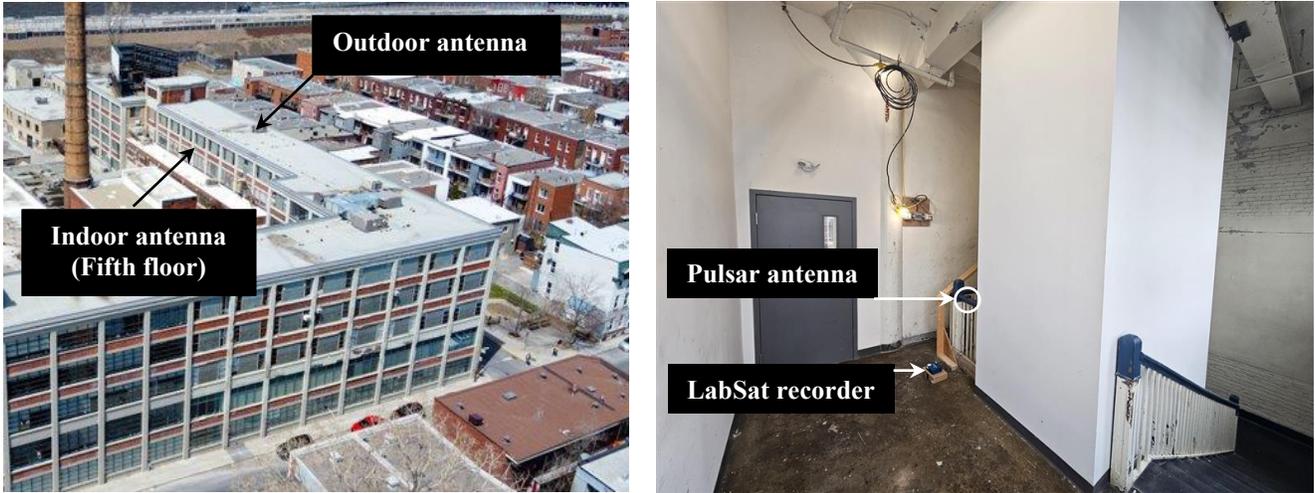

Figure 2: Outdoor and indoor data collection site

The received signal power levels for GPS and Pulsar signals from the outdoor antenna are compared in Figure 3 (a). As shown in the figure, Pulsar signals are about 10-15 dB stronger than GPS signals. $C/N_0$ for the Pulsar signal received indoors is compared against the signal received at the outdoor antenna in Figure 3 (b). As expected, indoor Pulsar signal is weaker compared to the outdoor signal and indoor antenna failed to acquire any GPS signals.

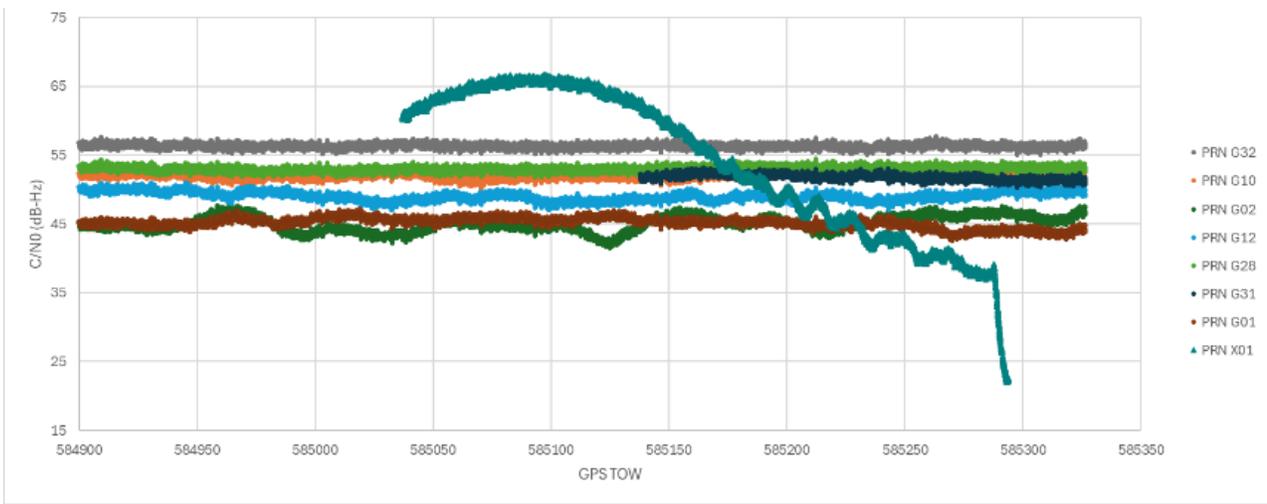

(a)

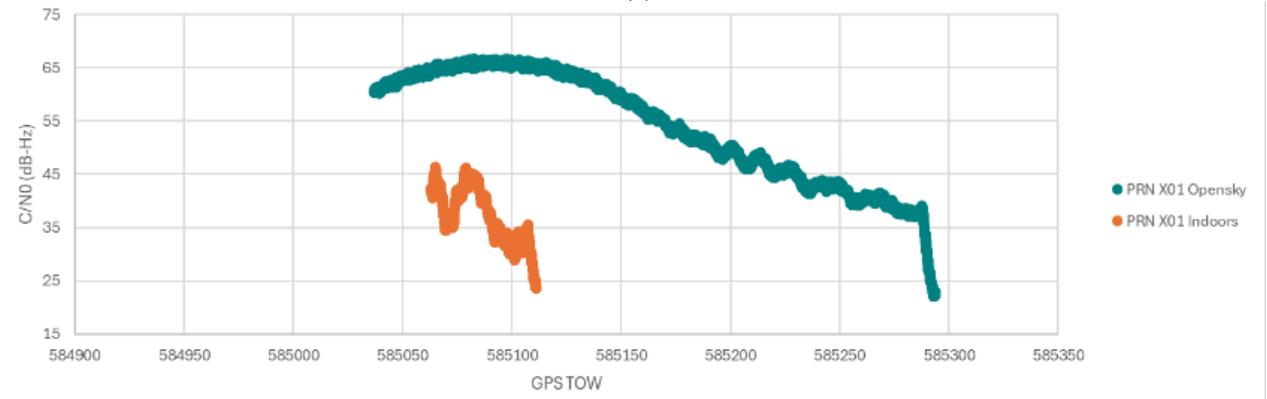

(b)

**Figure 3:** Comparison of received signal power comparison for (a) GPS outdoor vs. Pulsar outdoor (b) Pulsar outdoor vs. Pulsar indoor

This live-sky experiment demonstrated horizontal positioning accuracy at the sub-10-meter level under favorable carrier-to-noise conditions. The results confirm that leveraging LEO Doppler dynamics in conjunction with code measurements can deliver reliable positioning and timing continuity, paving the way for practical deployment in availability-critical applications. Indoor observations were available for just less than a minute; with these observations, position errors of less than 40 m were achieved. Position accuracy metrics for outdoor and indoor cases for the early live sky trials are given in Table 4.

Table 4: SSP accuracy for Pulsar-0 early live sky trials

| Scenario | Error [m] | | | | |
|---|---|---|---|---|---|
| | East | North | Up | Horizontal | 3D |
| Outdoor | 2.13 | 4.86 | 17.61 | 5.30 | 18.39 |
| Indoor | −3.82 | −38.51 | 93.31 | 38.70 | 101.02 |

In case of indoor data, in addition to weaker signals, there is poor geometry and measurements are noisy. As mentioned earlier, literature suggests that damping term in damped LM helps in solving ill-conditioned problems. LM method not only made the optimization numerically stable enough to converge tightly but also yielded sub-10 m position accuracy for several indoor logs as given in Table 5. For most of the nonlinear estimators, initial state of the estimator matters; initial state close to the user was chosen for these tests. The fact that LM converges tightly in these conditions indicates good internal consistency of the satellite states and observables from Pulsar.

Table 5: SSP accuracy for indoor logs from Pulsar-0 early live sky trials

| Date | Site | Horizontal error [m] |
|---|---|---|
| 27 Sep 2025 | 3rd Floor Window | 2.1 |
| 7 Nov 2025 | 5th Floor Stairwell | 2.0 |
| 17 Nov 2025 | 5th Floor Stairwell | 0.6 |
| 18 Nov 2025 | 5th Floor Stairwell | 1.1 |
| 24 Nov 2025 | 5th Floor Stairwell | 3.0 |

## 6. CONCLUSIONS

Xona is deploying Pulsar, a LEO commercial satellite navigation system designed to complement GNSS and deliver resilient PNT in constrained environments. LEO's higher received power and rapid geometry change enable SSP for stationary or semi-stationary users by stacking pseudorange and Doppler over a pass. This paper presented a GNSS-standard formulation for SSP, discussed algorithmic choices and modeling specifics important for LEO, and reported simulation and early live-sky findings. We demonstrated the feasibility of Pulsar SSP for availability-driven use cases by jointly exploiting pseudorange and Doppler over a pass and estimating receiver position with clock bias/drift. Preliminary live testing of Pulsar-0 signals demonstrates that positioning accuracies better than 10 m are achievable in both outdoor and indoor environments, indicating significant potential for reliable navigation under challenged conditions. Software simulations further show that combining observations across multiple passes significantly improves position accuracy relative to single-pass solutions.

## 7. USE-CASE OVERVIEW AND FUTURE WORK

### 7.1. Applicability of Single-Satellite Positioning

Single-satellite positioning is most suitable for scenarios where availability and continuity take precedence over sub-meter accuracy. While the achievable accuracy typically ranges from a few meters to tens of meters depending on pass duration and environmental conditions, this level of precision is adequate for many operational needs (Baron, Gurfil, & Rotstein, 2024). SSP is particularly valuable in IoT asset tracking for fixed or slow-moving objects, enabling zone-level localization, geofencing, and exception alerts with low-power receivers using only code and Doppler measurements (Baron, Gurfil, & Rotstein, 2024; Xu, Liu, Lei, Fang, & Jiang, 2025). It also supports logistics applications such as container and shipment tracking, providing yard or warehouse-level identification and coarse movement states during handling, with literature showing meter-class feasibility under benign conditions. Critical infrastructure deployments benefit from SSP through site-level localization and robust timing for distributed sensors, leveraging higher LEO signal power and authentication for resilience against interference. Furthermore, SSP excels in semi-indoor or urban canyon environments where MEO GNSS signals are attenuated, as LEO's stronger link budget and rapid Doppler dynamics maintain fixes over short arcs (Foreman-Campins, López-Salcedo, & Lohan, 2025). The method's effectiveness stems from strong Doppler observability, which constrains clock drift and combined with time-varying line-of-sight across a pass, enables position and clock estimation without external aiding.

### 7.2. Advantages of the Method

SSP offers several technical and operational benefits that make it attractive for availability-first applications. It works with a single satellite over short observation windows, eliminating the need for dense constellations or ground reference networks. The Doppler-rich geometry of LEO satellites provides large range-rate sensitivity, enabling accurate clock-drift estimation and stabilizing batch solutions. SSP also minimizes user complexity and power consumption by relying solely on code and Doppler measurements, avoiding carrier-phase ambiguities and remaining compatible with GNSS-like hardware and firmware paths. Its layered resilience allows seamless fusion with GNSS when available, improving accuracy and integrity while aiding acquisition and tracking in contested or attenuated environments (Oak, et al., 2024; Prieto-Cerdeira, Anghileri, Ries, & Giordano, 2023). Finally, SSP enhances security and integrity through authenticated, higher-power Pulsar LEO signals, mitigating spoofing and jamming risks compared to legacy open signals. These combined advantages position SSP as a practical solution for applications demanding robust availability rather than centimeter-level precision.

### 7.3. Enhancements with 2–3 Satellites in View

When two or three LEO satellites are concurrently visible, the geometry and conditioning of the positioning solution improve significantly compared to single-satellite scenarios. With two satellites, horizontal components are strengthened, while three satellites allow full 3D position and clock estimation per epoch without relying solely on long batch windows. This improved dimensionality reduces dilution of precision and accelerates convergence, enabling shorter batch windows and lower damping factors for iterative solvers such as Levenberg–Marquardt. Accuracy also benefits from this enhancement: short batches with independent line-of-sight vectors can achieve sub-5 m positioning under benign conditions, provided good carrier-to-noise ratios and high-fidelity modeling of effects such as atmospheric delays. Robustness improves as well, since redundancy enables residual-based outlier rejection and RAIM-like integrity checks, reducing sensitivity to ephemeris or velocity errors. The formulation for multi-satellite processing typically retains the state vector as in Equation (1) for position, clock bias, and drift, stacking pseudorange and Doppler measurements from all visible satellites per epoch, and optionally modeling per-satellite code biases when calibration residuals are observed.

### 7.4. Implementation Considerations

Achieving reliable performance requires careful attention to modeling and algorithmic details. High-fidelity geometry corrections are essential: transmit time iteration and Earth rotation (Sagnac) must be applied to both range and range rate, and relativistic or atmospheric terms included where relevant. Weighting strategies should reflect realistic measurement variances for pseudorange and Doppler. Finally, Doppler sensitivity makes satellite velocity accuracy paramount; ephemeris and velocity errors should be characterized through hardware-in-the-loop testing and live-sky cross-checks. These considerations ensure that the theoretical benefits of SSP and multi-satellite enhancements translate into practical, robust implementations.

### 7.5. Summary and Path to Future Work

Single-satellite positioning provides a compelling availability-first solution for IoT, logistics, and infrastructure applications, but its accuracy and robustness improve markedly with two or three satellites in view. These enhancements reduce convergence time, strengthen geometry, and enable integrity monitoring, paving the way for sub-5 m performance under favorable conditions. However, realizing these gains depends on rigorous modeling, robust estimation techniques, and careful handling of Doppler and ephemeris sensitivities. Future work would involve research into real-time implementations on low-SWaP receivers.

Possible extensions to the multi-pass solution and adapting for contested environments will include introducing adaptive weighting based on residual statistics or signal quality indicators like carrier-to-noise ratio and implementing robust estimation methods to mitigate outliers. Assessing sensitivity of positioning methods to elevation angle of the pass, number of samples used in position computation, initial state of the estimator, etc., can provide useful insights towards finetuning the methods. Further improvements could involve inclusion of X5 observables along with X1 or integrating additional satellites, either LEO or GNSS, for hybrid positioning solutions. These enhancements would make the multi-pass approach more resilient and accurate in real-world conditions.

**ACKNOWLEDGEMENTS**

We acknowledge the support of the Canadian Space Agency (CSA) 24STDPU43.